\documentclass[aps,prc,twocolumn,showpacs,groupedaddress]{revtex4-1}
\usepackage{epsfig}
\usepackage{dcolumn}   % needed for some tables
\usepackage{bm}        % for math
\usepackage{amssymb}

\begin{document}
\newcommand{\mybibitem}{\item}

%%%%%%%%%%%%%%%%%%%%%%%%%%%%%%%%%%%%%%%%
\title{Conformal Holography of Bulk Elliptic Flow and Heavy Quark
Quenching in  Relativistic Heavy Ion Collisions}

\author{Jorge Noronha$^1$, Miklos Gyulassy$^{1}$, and Giorgio Torrieri$^{2}$}
%\email{noronha@phys.columbia.edu}
\affiliation{$^1$Department of
Physics, Columbia University, 538 West 120$^{th}$ Street, New York,
NY 10027, USA\\
$^2$Frankfurt Institute for Advanced Studies, Frankfurt am Main, Germany}

\begin{abstract}
We show that the ``perfect fluid'' elliptic flow of the bulk hadrons
and the unexpectedly strong quenching of heavy quark jet fragments in Au+Au 
reactions at 200 AGeV 
can be simultaneously accounted for within leading order
AdS/CFT holography with a common large t'Hooft coupling 
$\lambda=g^2_{YM} N_c \sim 30$. In contrast, weakly coupled quark-gluon
plasma models have so far failed to describe the observed correlation between
these soft and hard observables even for couplings
  extrapolated to $\alpha_s\sim 0.5$. We show
that phenomenological applications of the classical trailing string
AdS/CFT holographic solution are furthermore 
remarkably robust to higher order curvature corrections
$\mathcal{O}(1/\lambda^{3/2})$ in type IIB supergravity theories,
as well as to worldsheet fluctuation corrections $\mathcal{O}(1/\lambda^{1/2})$ that were 
not considered previously. We emphasize the importance of  
future measurements at RHIC and LHC of the correlation
between {\em identified} charm and beauty quark hard ($p_T > 10-30$
GeV) jet
quenching observables and low transverse momenta ($p_T\le 1$ GeV) bulk
elliptic flow observables to further tests 
the limits of applicability of conformal holography to 
strongly coupled quark-gluon plasma.

\end{abstract}

%%%%%%%%%%%%%%%%%%%%%%%%%%%%%%%%%%%%%%%%%

\date{\today}  \maketitle

%\clearpage

%%%%%%%%%%%%%%%%%%%%%%%%%%%%%%%%%%%%%%
%%%%%%%%%%%%%%%%%%%%%%%%%%%%%%%%%%%%%%
%\section{Introduction}
%%%%%%%%%%%%%%%%%%%%%%%%%%%%%%%%%%%%%%
%%%%%%%%%%%%%%%%%%%%%%%%%%%%%%%%%%%%%%

\section{Introduction}

Two of the most remarkable experimental discoveries
at the Relativistic Heavy Ion Collider (RHIC) 
are the observations of a factor five quenching of (high transverse
momentum or high quark mass) jets and the nearly ``perfect
fluid'' elliptic flow of low transverse momentum hadrons 
\cite{RHICwhitepapers}. These observations and bulk multiplicity
production
systematics
 have been interpreted as
providing evidence for the formation of two new forms of QCD  matter
1) a locally equilibrated
strongly-coupled 
Quark-Gluon Plasma (sQGP) and 2) its non-Abelian classical 
Color Glass Condensate (CGC)
initial nuclear field source \cite{Gyulassy:2004zy,Kharzeev:2001yq}.

However, it has been a challenge to find a single consistent
theoretical framework to explain simultaneously both soft
(long wavelength)  and
hard (short wavelength) properties of the sQGP phenomena. Attempts to explain bulk collective flow \cite{Afanasiev:2009wq,Adcox:2002ms,v2data} based on perturbative
QCD parton transport approaches
\cite{Danielewicz:1984ww,coesterpaper,majumder,fochler}, require
 large coupling extrapolations $\alpha_s=g_{YM}^2/4\pi \rightarrow 0.6$,
which on the other hand overestimate of the opacity of the sQGP
to high transverse momentum jets. At moderate coupling $\alpha_s\sim 0.4$
the observed opacity for light quark jets can be well accounted for
but at the expense of too high viscosity and
underestimating 
the elliptic flow. The heavy quark jet quenching data pose an
especially difficult challenge for perturbative QCD energy loss models
\cite{Djordjevic:2005db}.

For large values of the QCD coupling, the
t'Hooft parameter $\lambda=g_{QCD}^2 N_c > 20$ (with $N_c=3$) may already be large
enough to validate string theory inspired classical AdS/CFT holographic models
\cite{maldacena,Gubser:1998nz} of the sQGP.
In this approximation, the viscosity to entropy ratio
is naturally small \cite{viscobound} and close
to its lower unitarity bound \cite{Danielewicz:1984ww,viscobound}.
Classical strings in the AdS background furthermore 
provide a dual holographic model of heavy quark drag 
with $dE/dx\propto \sqrt{\lambda} ET^2/M_Q$
that can easily account for the high
opacity of the sQGP to heavy quark jets \cite{drag}. 
The key question that we address in this paper is whether
there exists a single value of $\lambda$ that could  account
for both the bulk elliptic flow as well as the 
strong heavy quark
quenching simultaneously. We test at the same time the
consistency of the CGC initial field source geometry out of
which the sQGP forms. 
 Our conclusion is that conformal holography
can indeed account for the observed hard/soft correlation data 
with the CGC initial geometry with $\lambda\sim 30$
while classical Glauber
initial geometry is not consistent within present experimental
errors. 
We emphasize the importance
of testing future identified charm and bottom jet quenching
data systematics and their correlation with soft bulk flow observables.

\section{Strongly-Coupled $\mathcal{N}=4$
Supersymmetric Yang-Mills and Holography}

The planar limit of strongly-coupled $\mathcal{N}=4$
Supersymmetric Yang-Mills (SYM) theory provides a
powerful phenomenological model of the sQGP 
at zero baryonic chemical potential in the temperature range where the trace anomaly is small. Soft collective flow phenomena are controlled
by the system's entropy density, $s(T)$, and shear viscosity,
$\eta(T)$. In this conformal $c_s^2=d\ln T/d\ln s=1/3$ theory, 
the bulk viscosity
vanishes and the dimensionless ratio $\eta/s$ is independent
of temperature. While conformal invariance is broken in real world QCD,
the conformal symmetry of $\mathcal{N}=4$ SYM 
played an important role in the conjectured classical supergravity dual 
description \cite{maldacena} of this CFT on a AdS$_5\times S_5$ curved spacetime.
 In the $N_c,\,\lambda \to \infty$ limits, the compactification of
a 10d type IIB string theory on a 5d sphere of radius $L$ 
leads to a 5d Einstein action with a negative
cosmological constant
\begin{equation}
\mathcal{A}=\frac{1}{16\pi G_5}\int d^5 x
\sqrt{-G}\left(\mathcal{R}+\frac{12}{L^2}+\ldots\right)\,,
\label{gravityaction}
\end{equation}     
 The effective 5d gravitational coupling is taken as 
$G_5 \sim 1/N_c^2\ll 1$, and the $AdS_5$ metric $G_{\mu\nu}$ is found  to be
a stationary solution 
with curvature
 $\mathcal{R}=-12/L^2$.  The t'Hooft coupling in the gauge theory is
identified with $L^2/\alpha'$, where $\sqrt{\alpha'}={\ell}_s$ is the
fundamental 10d string
length. The $\alpha'$ expansion in the gravity dual description 
is mapped into a
series in $1/\sqrt{\lambda}$ in the gauge theory \cite{Gubser:1998nz}.
The extra $\cdots$ terms in Eq.\ (\ref{gravityaction}) 
correspond to the corrections to the leading order supergravity
action that appear when $L^2/\alpha'$ and/or $N_c$ are large but
finite \cite{Gubser:1998nz}. The leading ``stringy'' corrections to type
IIB string theory were found to come 
from terms $\sim\alpha'^3 \mathcal{R}^4$ to the
tree level effective action \cite{Gubser:1998nz} and they induce
contributions of order $1/\lambda^{3/2}$ to the quantities mentioned
above \cite{Gubser:1998nz,Buchel:2004di,Buchel1}. 

Breaking of supersymmetry on the gauge theory side can lead in the
gravity dual description to lower order 
quadratic or cubic curvature corrections.
A class of Gauss-Bonnet generalizations of the effective 5d
Einstein action was considered in 
\cite{Brigante:2007nu,Kats:2007mq,Buchel:2008vz}. 
These corrections are 
characterized by a second assumed small dimensionless parameter,
$\lambda_{GB}\sim 1/N_c$, which is
related to the
central charges $c$ and $a$ that characterize
 the conformal anomaly in curved spacetime of the
dual CFT as noted in Eq.\ (2.14) of Ref.\ \cite{Buchel:2008vz}. Varying
$\lambda_{GB}$ provides a parametric way to explore small 
deformations of the original ${\cal N}=4\; SU(N_c)$ SYM theory. 
Interest in possible Gauss-Bonnet deformations of conformal holography
arose when Kats and Petrov \cite{Kats:2007mq}
found that for ${\cal N}=2\; Sp(N_c)$ SYM, $\lambda_{GB}= 1/8N_c$, and
the
KSS viscosity bound \cite{viscobound}, $\eta/s\ge 1/4\pi$, could be violated
by 17\% for $N_c=3$, in such models. Ref. \cite{Buchel:2008vz} further
showed that a large class of other effective CFTs lead to similar $\lambda_{GB}\propto 1/N_c$ effects. In Refs.\ \cite{Brigante:2007nu,Hofman:2008ar}
it was found however that causality and positive energy flow
limit deformations to a narrow parametric range
$-7/36 <\lambda_{GB}<9/100$. We find below that RHIC hard/soft
correlation observables are in fact compatible with $\lambda_{GB}=0$.

Conformal holography refers to the 
predicted temperature independence of $s/s_{SB}$, the entropy density of a very 
strongly coupled SYM compared to its ideal Stefan Boltzmann limit,
and also of the viscosity to entropy density ratio.
The heavy quark jet relaxation rate, $1/\tau_Q$, is
controlled by $\mu_{Q}=\sqrt{\lambda}\,\pi
T^2/2M_{Q}$ for a heavy quark with mass $M_Q$ in a plasma of
temperature $T$ \cite{drag,drag1}. 
The relaxation time is related to the heavy quark energy loss per unit length
through $\tau_Q(\lambda)= -1/(d\log
p/dt)=-1/(d\log E/dx)$, where $p=M_{Q}\gamma v$ and $v=p/E$. 

Our analysis is based on the following remarkably simple
algebraic expressions relating three fundamental properties of large $N_c$, $\mathcal{N}=4$ conformal SYM plasmas 
at large t'Hooft coupling $\lambda$:
\begin{eqnarray}
\frac{s}{s_{SB}} & = & \frac{3}{4}\left(1+\frac{c_3}{8
    \lambda^{3/2}}\right),
\label{sbargeneral}\\
 \frac{\eta}{s} & = & \frac{1}{4\pi}\left(1+\frac{c_3}{\lambda^{3/2}}\right),
\label{etasgeneral}\\
 \tau_Q^{-1} &
= & \mu_{Q} \left(1 + \frac{\kappa_1}{\lambda^{1/2}}+\frac{c_3}{16\lambda^{3/2}} \right)
\label{dpdtgeneral}
\end{eqnarray}
where $c_3=15\zeta(3)\approx 18$, $\kappa_1\sim -1$ is a new
worldsheet fluctuation correction amplitude discussed below. The $\lambda^{-3/2}$ correction to the entropy density ratio in Eq.\ (\ref{sbargeneral}) was found in \cite{Gubser:1998nz} while the analogous correction to $\eta/s$ was obtained from \cite{Buchel1}. The finite t'Hooft coupling correction to the heavy quark energy loss, $c_3\,\lambda^{3/2}/16$, is a new result reported here (see the derivation below) and it is needed for a consistent application of the strongly-coupled $\mathcal{N}=4$ SYM model to heavy ion reactions, as will be argued below.

\section{Heavy Quark Jets in $\mathcal{N}=4$ SYM at large (but finite) t'Hooft Coupling}

The hard observable we consider here is heavy quark
energy loss. An object of mass $M_Q$ in the fundamental representation
of $SU(N_c)$ added to SYM is described holographically using a $D7$
brane \cite{Karch:2002sh} that ends at a radial direction $u_m \sim
M_Q$ away from the black brane horizon $u_h \sim T$. A heavy quark
moving with constant velocity $v$ is then described in terms of a
curved string in the radial direction which has the upper end attached
to the bottom of a D7 brane while the lower end remains attached to
the black brane \cite{drag,drag1}. When $N_c\to \infty$ and $\lambda\gg 1$ (but finite) the
string dynamics is embedded in a black brane curved background spacetime of the type
\begin{equation}
\ ds^2 = G_{00}(u)dt^2 + G_{xx}(u) d\vec{x}^{\,2}+G_{uu}(u)du^2
\label{generalmetric}
\end{equation}
where 
\begin{equation}
G_{00}(u)= -\frac{u^2}{L^2}\left( 1-\frac{u_h^4}{u^4}\right)\left(1+O(\alpha'^3)\right)
\label{generalmetric00}
\end{equation}
\begin{equation}
G_{uu}(u)= \frac{L^2}{u^2}\left( 1-\frac{u_h^4}{u^4}\right)^{-1}\left(1+O(\alpha'^3)\right)
\label{generalmetric00}
\end{equation}
and $G_{xx}=u^2/L^2\left(1+O(\alpha'^3)\right)$ (the finite $\alpha'$ corrections to the metric we use can be explicitly found in \cite{Gubser:1998nz}). 

The classical Nambu-Goto string action (with dilaton ommitted) is
\begin{equation}
 \mathcal{A}_{NG}=-\frac{1}{2\pi \alpha'}\int d^2\sigma \sqrt{-g}
\label{nambugotoaction}
\end{equation}
where 
\begin{equation}g=
  \det\,g_{ab}=G_{\mu\nu}\partial_{a}X^{\mu}\partial_{b}X^{\nu}
\label{nambugotoaction}
\end{equation} 
is
the induced worldsheet metric, $\sigma^{a}=(\tau,\sigma)$ are the
internal worldsheet coordinates, $G_{\mu\nu}(X)$ is the background metric,
 and $X^{\mu}=X^{\mu}(\tau,\sigma)$ is
the embedding of the string in spacetime. By restricting the analysis
to the classical string  the energy loss always becomes proportional to
$\sqrt{\lambda}$ at leading order
 because of the string tension $1/(2\pi \alpha')$ in the NG action.

The trailing string ansatz (where $\tau=t,\,\sigma=u$ and
$X^{\mu}(t,u)=(t,x_{0}+vt+\xi(u),0,0,u)$) is assumed to describe the asymptotic
behavior of a string whose endpoint (``the heavy quark") moves with velocity $v$ in the $x$ direction and located at a
fixed AdS radial coordinate near the boundary $u_m\gg u_h$ \cite{drag,drag1}.
The black brane horizon coordinate $u_h \propto T \alpha'$ is
determined by $G_{00}(u_h)=0$. Using the ansatz above and the string's classical
equations of motion, it can be shown \cite{drag} that the drag force $dp/dt= - C
v/(2\pi \alpha')$, where $C$ is a constant determined by the
negativity condition that
\begin{equation}
\ g(u) = G_{uu}\left(G_{00}+v^2 G_{xx}\right)\left(1+\frac{C^2 v^2}{G_{00}G_{xx}}\right)^{-1}
<0
\label{inducedmetric}
\end{equation}
for $u_h \leq u \leq u_m$. However, both the numerator and
denominator in Eq.\ (\ref{inducedmetric}) change their sign
simultaneously at a certain $u^*$ \cite{drag} given by the root of
the  equation
$%\begin{equation}
\ G_{00}(u^*) +v^2 G_{xx}(u^*)=0
%\label{condition1}
$. This fixes $C=G_{xx}(u^*)$ and %or, equivalently,
$dp/dt=-v
\,G_{xx}(u^*)/(2\pi \alpha')$. Neglecting higher-order derivative
corrections in $\mathcal{N}=4$ SYM one finds $u^*=u_h \sqrt{\gamma}$,
where $\gamma=1/\sqrt{1-v^2}$. The condition that $u^*\leq u_m$
leads to a maximum ``speed limit'' for the heavy quark jet to be
consistent with this trailing string ansatz given by $\gamma_{\rm
  max}\leq u_m^2/u_h^2$ \cite{speedlimit}.

Using the metric derived in \cite{Gubser:1998nz} to $\mathcal{O}(\alpha\,'^{3})$ and the classical NG action, one can compute the effects of quartic corrections on the classical drag force and determine $u^*$ perturbatively to $\mathcal{O}(\lambda^{-3/2})$ as 
$ %\begin{equation}
 u^* =u_h \sqrt{\gamma} \left[1+ \frac{15}{32}\frac{\zeta(3)\, v^2}{\lambda^{3/2}}\left(5+\frac{5}{\gamma^2}-\frac{3}{\gamma^4}\right)\right]
%\label{newu*}
$%\end{equation}
and the drag force
\begin{equation}
\frac{dp}{dt} = -\sqrt{\lambda}\,T^2\frac{\pi}{2} v\gamma \left[1+\frac{15}{16}\frac{\zeta(3)}{\lambda^{3/2}}\left(1-\frac{197}{24\gamma^4}+\frac{67}{24\gamma^6}\right)\right] .
\label{newdpdt1}
\end{equation}
The analytical result for the classical heavy quark drag force at large (but finite) t'Hooft coupling in $\mathcal{N}=4$ SYM computed above is one of the main results of this paper. The heavy quark mass at $T=0$ is $M_Q=u_m/(2\pi \alpha')$ and, to leading order in $1/\lambda$, $u_m^2/u_h^2\simeq \frac{4M_Q^2}{\lambda T^2}$. Thus, the corrected $u^*$ displayed above defines a new speed limit $\gamma_m \simeq \frac{4M_Q^2}{\lambda T^2}\left[1-\frac{5}{16}\left(\frac{4\pi\eta}{s}-1\right)\right]$, after neglecting terms of $\mathcal{O}(1/\gamma,1/N_c)$. Note that $\gamma_m$ and $dp/dt$ decrease with increasing $\eta/s$.

The drag force experienced by a heavy quark in the strongly-coupled $\mathcal{N}=4$ SYM plasma can also receive finite $\lambda$ corrections from quantum fluctuations of the string worldsheet, as it will be discussed below.   

\subsection{Finite t'Hooft coupling corrections to heavy quark energy loss}

While the large amount of supersymmetries in
$\mathcal{N}=4$ SYM excludes the corrections to $\eta/s$ due to
$\mathcal{R}^2$ and $\mathcal{R}^3$-like terms in its 10d gravity dual action
\cite{Gubser:1998nz}, this is not the case for observables such as the string drag
or  the effective heavy quark potential
\cite{wilson,Hou:2008pg,Chu:2009qt}.
 There are no known reasons to exclude terms in the
$\alpha'$ expansion of the 2d world sheet theory.
Physically, these corrections correspond to fluctuations of
the string worldsheet around its minimum area. For instance, the
$\alpha'$ corrections just from the worldsheet string loops defined on
top of a supergravity background (where the gravity dual's action
contain only quadratic terms in the spacetime derivatives) would
contribute to the energy loss $E$ of a heavy quark in the medium as
follows
\begin{eqnarray}
\frac{d \ln E}{d\hat{x}} &=& -\frac{\sqrt{\lambda}}{2}\,\frac{T}{M_Q}v
\left(1+\frac{\kappa_1(v)}{\sqrt{\lambda}}+\frac{\kappa_2(v)}{\lambda}+\frac{\kappa_3(v)}{\lambda^{3/2}}
\right. \nonumber \\ &+&
\left. \frac{\kappa_4(v)}{\lambda^2}+\mathcal{O}(\lambda^{-5/2})
\right)
\label{loopexpansion}
\end{eqnarray} 
where $\hat{x}\equiv (\pi T) x$ is a dimensionless quantity, $E\equiv
M_Q \gamma$, and $\kappa_i$ with $i=1,\ldots,4$ denote the
contribution from $i-th$ string loop corrections. The first term
outside the parenthesis is the standard result obtained before in the
$\lambda \to \infty$ limit \cite{drag,drag1}. For physical quark
masses the classical string description is only applicable when
$\frac{\sqrt{\lambda}}{2}\,\frac{T}{M_Q} < 1$, which
indicates that $\lambda$ cannot indeed be taken to be infinitely
large. In a critical string theory such as type IIB, the loop
coefficients are expected to be finite and, in general, $v$
dependent. The 1-loop coefficient of the correction to the heavy quark
potential in the vacuum (which gives a term analogous to the
$\kappa_1/\sqrt{\lambda}$ above) was recently found numerically to be
$\sim -1.33$ \cite{Chu:2009qt}.

As shown in Eq.\ (\ref{newdpdt1}), the $\alpha'^3$ correction to the background studied in \cite{Gubser:1998nz} affects the energy loss. This contribution can be taken into account independently of the worldsheet fluctuations at least to lowest order. The ultrarelativistic limit of Eq.\ (\ref{newdpdt1}) is
\begin{equation}
\frac{d\ln E}{d\hat{x}} = -\frac{\sqrt{\lambda}}{2}\,\frac{T}{M_Q} 
\left[1+\frac{15}{16}\frac{\zeta(3)}{\lambda^{3/2}}\right] .
\label{newdpdt}
\end{equation}
This is the new term that was included in Eq.\ (\ref{dpdtgeneral}). Note that with this type of correction suggests that the magnitude of
the energy loss becomes larger than the value obtained at infinite
coupling as $\lambda$ decreases (or $\eta/s$ increases), which is
certainly counterintuitive and most likely unphysical. This poses no
harm if after the string loop corrections in
Eq.\ (\ref{loopexpansion}) are included the correct physical behavior
(i.e., smaller coupling means smaller energy loss) is recovered. This
implies that worldsheet fluctuations must be taken into account in the
study of finite $\lambda$ corrections to the energy loss in
$\mathcal{N}=4$ SYM.

If one includes both the string loop corrections and the bulk-induced
new term one obtains (assuming that the $v\to 1$ limit of the loop
corrections is well defined)

\begin{eqnarray}
\frac{d \ln E}{d\hat{x}} &=& -\frac{\sqrt{\lambda}}{2}\,\frac{T}{M_Q}
\left[1+\frac{\kappa_1(1)}{\sqrt{\lambda}}+\frac{\kappa_2(1)}{\lambda}+
\frac{1}{\lambda^{3/2}}\right. \nonumber \\
  & & \left. \left(\frac{15}{16}\zeta(3)+\kappa_3(1)\right) 
+\frac{\kappa_4(1)}{\lambda^2}+\ldots \right] 
\label{loopexpansion1}
\end{eqnarray} 
%\end{document}
A comparison between Eqs.\ (\ref{sbargeneral}), (\ref{etasgeneral}),
and (\ref{loopexpansion1}) shows that the heavy quark energy loss can
be in principle much more sensitive to finite $\lambda$ corrections
than the bulk quantities. This is also true, for instance, for heavy
quark bound states described within AdS/CFT \cite{Noronha:2009ia}. On
the other hand, bulk quantities such as $\eta/s$ are finite at any
(large) value of the coupling.

For our applications,  we consider the range $\lambda\sim 5-30$ and $N_c=3$. Also, we neglect other formally higher order
corrections \cite{Buchel1} $\sim\mathcal{O}(\sqrt{\lambda}/N_c^2)$ to the $\mathcal{N}=4$ SYM entropy density and shear viscosity and set $\kappa_2=\kappa_3=\kappa_4=0$ in this first attempt to test conformal holography of hard and soft observable correlations in high energy A+A collisions.

\section{Correlation between the heavy quark jet $R_{AA}$ and bulk elliptic flow}

The dependence of soft phenomena on $\eta/s$ in nearly perfect fluids
can be studied using viscous relativistic hydrodynamics
\cite{Luzum:2008cw}. Once $v_2$ as a function of $\eta/s$ is known
from hydrodynamic simulations, one can use Eq. (\ref{etasgeneral}) to
obtain its dependence on $\lambda$ and estimate how much elliptic flow
is generated in a strongly-coupled SYM plasma. In order to compute
$v_2(p_T,\eta/s)$ for a given centrality interval $\mathcal{C}$, we
employ a linear fit to the numerical results of \cite{Luzum:2008cw}
for both Glauber \cite{glauber} and CGC initial transverse profiles \cite{Drescher:2006pi,CGCinitial}
\begin{equation}
v_2(p_T,\eta/s,\mathcal{C})= a(p_T)\, \epsilon_2(\mathcal{C}) \,( 1- a_1\,\eta/s)
\end{equation}
where $\epsilon_2(\mathcal{C})=\langle y^2-x^2\rangle_{\mathcal{C}}
/\langle x^2+y^2\rangle_\mathcal{C}$ is the average initial elliptic
geometric eccentricity for the centrality class $\mathcal{C}$. We
consider the 20-60\% centrality class because, as shown in Fig.\ 23 of
\cite{Afanasiev:2009wq}, there is good agreement at $p_T\sim 1$ GeV
between STAR $v_2(4)$ and PHENIX $v_2(BBC)$ data and non-flow effects
\cite{poszkanzer} are reduced. To rescale the minimum bias results of
Ref.\ \cite{Luzum:2008cw} to the considered 20-60\% centrality class
we use the factor
$\epsilon^{Glaub}_2(20-60\%)/\epsilon^{Glaub}_2(0-92\%)=0.317/0.281=1.128
$ from Ref.\ \cite{Adcox:2002ms}. Our fit to the rescaled numerical
results of \cite{Luzum:2008cw} gives $a_1\approx 2.5$ and
$a(p_T=1)\epsilon_2(20-60\%)\approx 0.14 \;(0.098)$ for CGC (Glauber)
initial conditions. In this case, the Glauber initial conditions do not generate enough elliptic flow and only the CGC initial conditions can fully describe the elliptic flow data.

The nuclear modification factor of single non-photonic electrons,
$R_{AA}^e(p_T)$, obtained from quenched heavy quark jets can be
computed at strong coupling \cite{Horowitz:2007su} by using the
generalized energy loss in Eq.\ (\ref{loopexpansion1}) to compute the
path length dependent heavy quark fractional energy loss
$\epsilon$. In this case $R_{AA}=\langle(1-\epsilon)^{n_Q}\rangle$,
where $n_{Q}(p_T)$ is the flavor dependent spectral index
$n_{Q}+1=-\frac{d}{d\,\ln
  p_T}\ln\left(\frac{d\sigma_{Q}}{dydp_T}\right)$ obtained from FONLL
production cross sections \cite{cacciari}. The path length average of
the nuclear modification at impact parameter $b$ is computed using a
Woods-Saxon nuclear density profile with Glauber profiles
$T_A(\vec{x}_{\perp})$ with $\sigma_{NN}=42$ mb. For 0-10\% centrality
triggered data both Glauber and CGC geometries lead to similar
numerical results for $R_{AA}$ \cite{Drescher:2006pi}. The
distribution of initial hard jet production points at a given
$\vec{x}_{\perp}$ and azimuthal direction $\phi$ is taken to be
proportional to the binary parton collision density,
$T_{AA}(\vec{x}_{\perp},b)$.

We consider a longitudinally expanding local (participant) parton
density $\rho(\vec{x}_{\perp},b)=\chi
\rho_{part}(\vec{x}_{\perp},b)/\tau$, where $\chi\equiv
(dN_{\pi}/dy)/N_{\rm part}$ and $\rho_{part}$ is the Glauber
participant nucleon profile density. In our calculations we use a
reduced temperature $T_{CFT}=0.74 (S/S_{SB})^{1/3}T_{QCD}$ to take
into account the fewer number of degrees of freedom in a
strongly-coupled QCD plasma, which is similar to the prescription
given in \cite{Gubser:2006qh}. The heavy quark modification factor is
\begin{eqnarray}
\ R_{AA}^{Q}(p_T,b)&=&\int_{0}^{2\pi}d\phi\int
d^2\vec{x}_{\perp}\frac{T_{AA}(\vec{x}_{\perp},b)}{2\pi\, N_{\rm
    bin}(b)} \nonumber \\ &\;& \hspace{-1 in} \times
%\exp\left[-n_{Q}(p_T)F_Q(\vec{x}_{\perp},\phi)\right]
\exp\left[-n_{Q}(p_T)\int_{\tau_0}^{\tau_f}
  \frac{d\tau}{\tau_Q(\vec{x}_{\perp}+\tau \hat{e}(\phi),\phi)}\right]
\label{Raaq}
\end{eqnarray}
where $N_{\rm bin}$ is the number of binary collisions and the
relaxation time $\tau_Q$ is defined in terms of
(\ref{loopexpansion1}).

Here, $\tau_0=1$ fm/c is the assumed plasma
equilibration time and $\tau_f$ is determined from
$T(\vec{\ell},\tau_f)=T_f=140$ MeV, i.e, the time at which the local
temperature falls below a freeze-out temperature taken from
\cite{Luzum:2008cw}. We assume that the non-photonic electron
modification factor is $R_{AA}^e(p_T)=0.6 R_{AA}^{bottom}(p_T^Q)+0.4
R_{AA}^{charm}(p_T^Q)$, where $p_T^Q=p_T/0.7$ is an estimate of
fragmentation effects \cite{Djordjevic:2005db}.

\begin{figure}[ht]
\epsfig{file=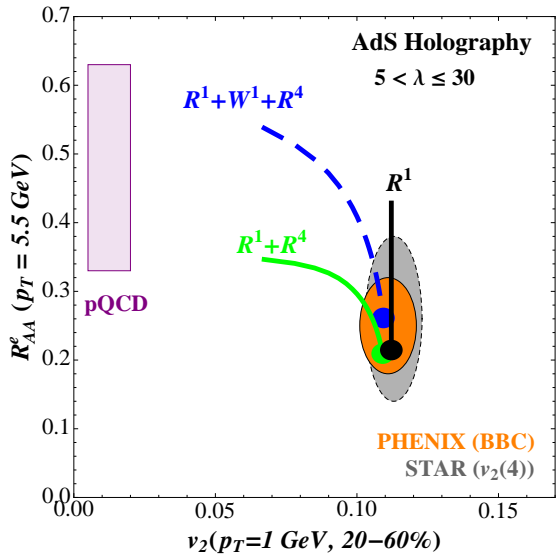,width=3in,clip=}
\caption{\label{Raaxv2plot} (Color online) The observed 
  correlation between the nuclear modification,
  $R_{AA}^e(p_T=5.5)$, of  0-10\% centrality non-photonic electrons
  from heavy quark jets and the bulk elliptic flow moment
  $v_2(p_t=1\;{\rm GeV})$ of pions
  produced in 20-60\% semi-central reactions is shown
by the  orange (grey) ellipse
  from  PHENIX (STAR) measurements of Au+Au
  200 AGeV data at RHIC
  \cite{Afanasiev:2009wq,Adler:2005xv,starelectronRaa,v2data}.
Curves show predictions based on conformal holography with
t'Hooft coupling in the range $5\le\lambda\le 30$ 
and assuming the initial participant geometry predicted by the KLN Color
Glass Model \cite{Kharzeev:2001yq,Drescher:2006pi,CGCinitial} 
and with  total entropy constrained by the measured  
total pion multiplicity $dN_{\pi}/dy=1000$ in central collisions. 
 The   horizontal black line corresponds to the lowest order supergravity result
 with $(\kappa_1=0,c_3=0)$ in Eqs.\ (2-4).   The green curve is for
  $(\kappa_1=0,c_3\approx 18)$ including only ${\cal R}^4$ stringy
 curvature corrections to   the ${\cal R}^1$ Einstein action.
 The dashed blue curve includes, in addition, an estimate 
for the leading worldsheet
 fluctuation correction $W^1$ with $\kappa_1=-1.33$ (see text). 
The purple band denotes a range of these 
observables far from the observed data as 
computed in pQCD-based transport models 
\cite{coesterpaper,Djordjevic:2005db}.}
\end{figure}
%\end{center}
%\vspace{-0.35in}

The $\mathcal{N}=4$ SYM prediction for the soft $v_2$ and hard $R_{AA}^e$ correlation, computed to lowest order in the coupling (supergravity limit), is shown in Fig.\ 1 (horizontal, red line) for $\lambda =
5,\ldots,30$. Only the results obtained with CGC initial conditions
are shown. The eccentricity of the Glauber initial geometry models are
too low to fit the data. In this case $4\pi\eta/s=1$ 
and the energy loss is $\sim
\sqrt{\lambda}$ \cite{drag,drag1}. The agreement between this lowest
order calculation and the RHIC data is rather remarkable for large
$\lambda\sim 20-30$. While finite
coupling corrections to $\eta/s$ only slightly shift $v_2$ to the
left, $R_{AA}^e$ involves an exponential of the energy loss and, thus,
one may expect this quantity to display a stronger sensitivity to
$1/\lambda^n$ corrections. While most of the coefficients in the
expansion (\ref{loopexpansion1}) are unknown, a useful check of the
robustness of the lowest order result can be done by repeating the
calculations with the known $1/\lambda^{3/2}$ corrections to $\eta/s$
in Eq.\ (\ref{etasgeneral}) and approximating Eq.\ (\ref{loopexpansion1}) with
the 1-loop worldsheet correction coefficient to the heavy quark potential in vacuum $\kappa_1=-4/3$ found in \cite{Chu:2009qt} (with other
corrections neglected). This leads to the dashed blue curve in
Fig.\ 1. Note that because $\lambda\sim 30$ is so large even
this formally much larger correction is negligible compared to the
leading supergravity prediction. We have checked \cite{Noronha:2009vz}
 that adding phenomenologically a 
 ``perturbative" quadratic Gauss-Bonnet 
curvature corrections to the Einstein action dual to $\mathcal{N}=4$
SYM \cite{Brigante:2007nu,Kats:2007mq,Buchel:2008vz} including also
its modification
of the heavy quark energy loss from \cite{poritz,Fadafan:2008gb},
following the suggestion made in Ref.\ \cite{Buchel:2008vz}, does not
lead to any significant changes in the results shown in Fig.\ 1 for
the range $|\lambda_{GB}| < 0.09$ allowed by causality and positivity
of energy flux. This indicates that possible $1/N_c$ corrections do not alter
the conclusions of this paper.

The light purple band near the top right corner of Fig.\ 1 
shows the range of hard/soft correlations that far from the data
predicted by pQCD based transport models
\cite{coesterpaper,Djordjevic:2005db}. 
Our results suggest  that the
sQGP discovered at RHIC are remarkably well described 
at both large and small wavelengths by an approximate conformal
holographic model, while the observed hard/soft correlation
 severely challenges perturbative QCD 
approximations. 

We conclude by demonstrating that future identified charm and 
beauty jet quenching data at RHIC and LHC will soon allow much more
stringent and sensitive test of  holographic QCD phenomenology. 
\begin{figure}[ht]
\epsfig{file=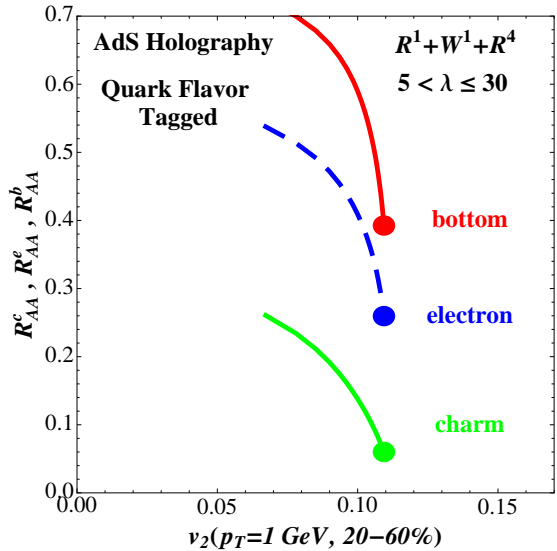,width=3in,clip=}
\caption{\label{RaabcWR4} (Color online) The predicted correlation
between identified heavy quark jet quenching, $R_{AA}^Q$, and 
soft elliptic flow, $v_2(p_T=1\;{\rm GeV})$, based on approximate $R^1+W^1+R^4$ 
conformal holography that
includes leading order worldsheet fluctuations and quartic curvature
corrections via Eqs.\ (2-4). The upper red curve corresponds to bottom
quark flavor tagged jets while the lower green curve corresponds to charm
flavor tagged jets. The blue dashed curve is the nuclear modification 
of single electrons from b and c jets estimated from $R_{AA}^e\approx 0.6
R_{AA}^b+0.4 R_{AA}^c$ with $p_T^e\approx 0.7 p_T^Q$, as also shown in Fig.\ 1.}
\end{figure}
%\end{center}
%\vspace{-0.35in}
The predicted small double ratio $R_{AA}^c/R_{AA}^b\ll 1$ is a robust
signature of AdS string drag models as emphasized in 
\cite{Horowitz:2007su}, which  differs
significantly from pQCD-based transport models and can be readily
tested once flavor tagged jet measurement in A+A reactions become
feasible at RHIC and LHC.

An important open theoretical problem is to generalize the above
analysis to non-conformal holographic models that 
that take into account the conformal anomaly near $T_c\sim 170$ MeV 
as predicted by lattice QCD.
Furthermore a more consistent and
quantitative light quark/gluon jet non-conformal holographic 
QCD phenomenology will need to be developed
to account simulaneously also for the observed nuclear modification
high $p_T$ pions from light quark and gluon jets, $R_{AA}^\pi(p_T)$.
The holographic string drag model postulated for heavy
quarks is inapplicable for light quark jets
 and may break down already for charm quark jets. 
This possibility underlines the
 need for the next generation experiments at RHIC and LHC that can
 measure simultaneously  charm and bottom jet quenching observables
 as well as pion nuclear modification factors
and their correlations with the bulk elliptic flow and total
entropy production.

We thank A.\ Dumitru, S.\ Gubser, W.\ Horowitz, A. Poszkanzer,
B.~Cole, R.~Lacey, and W.~Zajc for useful comments. J.N. and
M.G. acknowledge support from US-DOE Nuclear Science Grant
No. DE-FG02-93ER40764. G.T. acknowledges the financial support received from the Helmholtz International
Center for FAIR within the framework of the LOEWE program
(Landesoffensive zur Entwicklung Wissenschaftlich-\"Okonomischer
Exzellenz) launched by the State of Hesse.

\end{document}